\documentclass{jfm}

\usepackage{graphicx}
\usepackage{natbib}
\usepackage{textcomp}
\usepackage{mathtools}
\usepackage{stmaryrd} 
\usepackage{makecell}
\usepackage{booktabs}
\usepackage{leftidx}
\usepackage{tikz}
\usetikzlibrary{arrows,calc}
\usepackage{pgfplots}
\pgfplotsset{compat=1.3}
\usepackage{pifont}
\usepackage{xspace}
\usepackage{microtype}
\ifCUPmtlplainloaded \else
  \checkfont{eurm10}
  \iffontfound
    \IfFileExists{upmath.sty}
      {\typeout{^^JFound AMS Euler Roman fonts on the system,
                   using the 'upmath' package.^^J}%
       \usepackage{upmath}}
      {\typeout{^^JFound AMS Euler Roman fonts on the system, but you
                   dont seem to have the}%
       \typeout{'upmath' package installed. JFM.cls can take advantage
                 of these fonts,^^Jif you use 'upmath' package.^^J}%
      }
  \else
  \fi
\fi

\ifCUPmtlplainloaded \else
  \checkfont{msam10}
  \iffontfound
    \IfFileExists{amssymb.sty}
      {\typeout{^^JFound AMS Symbol fonts on the system, using the
                'amssymb' package.^^J}%
       \usepackage{amssymb}%
         \let\leq=\leqslant
         \let\geq=\geqslant
      }{}
  \fi
\fi

\ifCUPmtlplainloaded \else
  \IfFileExists{amsbsy.sty}
    {\typeout{^^JFound the 'amsbsy' package on the system, using it.^^J}%
     \usepackage{amsbsy}}
    {\providecommand\boldsymbol[1]{\mbox{\boldmath $##1$}}}
\fi

\newcommand\Rey{\mbox{\textit{Re}}}  
\newcommand\Ca{\mbox{\textit{Ca}}}  
\newcommand\Bo{\mbox{\textit{Bo}}}  

\newsavebox{\astrutbox}
\sbox{\astrutbox}{\rule[-5pt]{0pt}{20pt}}

\newcommand{\bild}[1]{figure~\ref{#1}}

\newcommand{\gleich}[1]{equation~\eqref{#1}}

\newcommand{\definiert}{\mathrel{\mathop:}=}
\newcommand{\definiertb}{=\mathrel{\mathop:}}
\newcommand{\vek}[1]{\boldsymbol{#1}}
\newcommand{\vektilde}[1]{\boldsymbol{\tilde{#1}}}
\newcommand{\sprung}[1]{\left\llbracket #1 \right\rrbracket}
\newcommand{\partiell}[2]{\frac{\partial #1}{\partial #2}}
\newcommand{\ord}[1]{^{(#1)}\xspace}
\newcommand{\landauca}[1]{\mathcal{O}(\Ca^{#1})}
\newcommand{\landaueps}[1]{\mathcal{O}(\varepsilon^{#1})}

\newcommand{\kYn}{\leftidx_k Y_n\ord1}

\makeatletter

\newcommand{\Rmnum}[1]{{\expandafter\@slowromancap\romannumeral #1@}}
\makeatother

\title[]{Driven particles at fluid interfaces acting as capillary dipoles}

\author[Aaron D\"orr and Steffen Hardt]%
{Aaron D\"orr and Steffen Hardt}

\affiliation{Institute for Nano- and Microfluidics, Center of Smart Interfaces, Technische Universit\"at Darmstadt, Alarich-Weiss-Stra\ss e 10, 64287 Darmstadt, Germany}

\pubyear{2015}
\volume{770}
\pagerange{5--26}
\date{\today}

\begin{document}

\maketitle

\begin{abstract}
The dynamics of spherical particles driven along an interface between two immiscible fluids is investigated asymptotically. Under the assumptions of a pinned three-phase contact line and very different viscosities of the two fluids, a particle assumes a tilted orientation. As it moves, it causes a deformation of the fluid interface which is also computed. The case of two interacting driven particles is studied via the Linear Superposition Approximation. It is shown that the capillary interaction force resulting from the particle motion is dipolar in terms of the azimuthal angle and decays with the fifth power of the inter-particle separation, similar to a capillary quadrupole originating from undulations of the three-phase contact line. The dipolar interaction is demonstrated to exceed the quadrupolar interaction at moderate particle velocities.
\end{abstract}


\section{Introduction}
Macroscopic phenomena involving particles attached to the interface between two immiscible fluids offer a wide variety of interesting fields of research \citep[e.g.][]{Binks2002,Ghezzi2001,Kralchevsky2000,Koser2013,Horozov2005}. In contrast to bulk suspensions \citep{Russel1989}, both the fluid interface itself as well as the difference in the properties of the two fluids, on the one hand, influence the forces acting on individual particles, and, on the other hand, introduce additional mechanisms for particle-particle interactions.

Regarding individual particles, researchers have studied the particles' equilibrium configuration at the fluid interface \citep{Nicolson1949,Huh1974,Singh2004}, their temporal response to external disturbances \citep{Singh2011}, the work necessary to detach particles from a fluid interface \citep{Pitois2002}, their interfacial mobility under external forcing \citep{Fulford1986,O'Neill1986,Petkov1995,Danov1995,Danov1998,Cichocki2004,Fischer2006,Pozrikidis2007,Ally2010,Bawzdziewicz2010} as well as the related diffusion coefficient \citep{Radoev1992,Chen2008,Du2012,Sriram2012}. Unexpected behaviour in the diffusion coefficient has been observed \citep{Chen2008,Sriram2012} and attributed to possible configurational changes in the system, e.g. detachment of particles from the fluid interface.

Particle-particle interactions arising from deformations of the fluid-fluid interface due to external forces and/or irregular menisci have been studied extensively \citep{Nicolson1949,Chan1981,Fortes1982,Kralchevsky2000,Stamou2000,Ghezzi2001,Binks2002,Singh2005,Danov2005,Vassileva2005,Oettel2005,Dominguez2008,Oettel2008}. In particular, \citet{Vassileva2005} and \citet{Singh2005} have discussed hydrodynamic interactions of interfacial particles approaching each other, the latter reference containing direct numerical simulations of interacting floating particles. Concerning capillary interactions at large separations, a modelling approach is the Linear Superposition Approximation or LSA, introduced by \citet{Nicolson1949} and occasionally used within later studies \citep[e.g.][]{Chan1981,Vassileva2005,Oettel2005}. The LSA exploits knowledge of the fluid-fluid interfacial deformation around a single particle by approximating the two-particle deformation via linear superposition of two single-particle deformations.

As already mentioned, a number of studies have dealt with single driven particles at fluid-fluid interfaces. Those studies may be divided into two groups. The first group assume the fluid-fluid interface being either planar \citep{O'Neill1986,Danov1995,Cichocki2004,Fischer2006,Bawzdziewicz2010,Ally2010,Wuerger2014} or having a given shape \citep{Petkov1995} unaffected by the particle motion, thus focusing on a purely hydrodynamic problem and aiming at the mobility tensor of the particle. A part of the publications from the first group consider interfacial incompressibility or viscosity, properties originating from the presence of surfactants or associated with membranes, as opposed to pure interfaces. The second group investigate the deformation of a fluid interface caused by a moving particle. This group also includes works on particles moving parallel to the fluid interface without touching the latter \citep{Lee1979,Berdan1982}. In all of these studies, the interfacial shape approximated either asymptotically \citep{Lee1979,Berdan1982,Radoev1992}, numerically \citep{Fulford1986,Pozrikidis2007}, or based on a combined approach \citep{Danov1998}, shows maximum deviation from the planar interface on the axis of the particle motion and zero deviation along the axis perpendicular to it, corresponding to a dipolar deformation.

Remarkably, among the studies on driven particles, only the methods of \citet{Pozrikidis2007}, and, in the context of floating particles, of \citet{Singh2005} require the specification of a boundary condition for the fluid-fluid interfacial shape at the particle surface. In both cases, a constant contact angle is specified. By contrast, when a particle moves along an interface with a pinned three-phase contact line, which is a physically reasonable assumption for relatively small particle velocities, the contact angle varies along the three-phase contact line as a result of hydrodynamic forces acting on the fluid interface. Therefore, we use the assumptions of a pinned three-phase contact line and a varying contact angle throughout the present paper. The effect of a limited range of contact angle hysteresis will be discussed further below.

As an additional motivation for the present study, we consider a particle driven by an external force acting on the particle's centre-of-mass and parallel to the fluid interface. As the particle moves subject to the force, it approaches a steady-state velocity until finally the drag force balances the external force. However, when there exists a viscosity difference between the two fluids, the drag force's line of action generally does not run through the particle's centre-of-mass. Therefore, the drag force is accompanied by a torque on the particle. As a consequence, the particle is rotated until the hydrodynamic torque is balanced by the torque due to interfacial tension. Thus, the initial particle configuration is not sustained because it is inconsistent with the balance of angular momentum. To the best of our knowledge, the effect just described has not been studied before in the context of interfacial particles and shall be the main topic of the present work. As a side remark, a similar configurational change is known from ship hydrodynamics \citep{Rawson2001}. In that context, the configuration is quantified in terms of the so-called trim angle. We adopt that notation for the present paper.

In the following, we describe the interplay between hydrodynamic and interfacial tension forces by means of an asymptotic model. The flow field is calculated as a first step and then utilized to approximate the interfacial shape. In the last step, the angular momentum balance for the particle yields the particle trim angle. Once the final configuration of the one-particle system is known, the LSA enables us to discuss the resulting dipolar particle-particle interactions. Finally, we comment on possible consequences and extensions of the model.

\section{Modelling approach}

We consider a rigid spherical particle of radius~$a$ attached to the interface~$\Sigma_{12}$ between two immiscible fluids~1 and~2 as depicted in \bild{fig:Kugel_GGW}.
\begin{figure}
\centering%
\def\breiteb{5cm}
\begin{tikzpicture}[x=0.1*\breiteb,y=0.1*\breiteb]
\node[anchor=south west,inner sep=0pt] at (0,0) {\includegraphics[height=\breiteb]{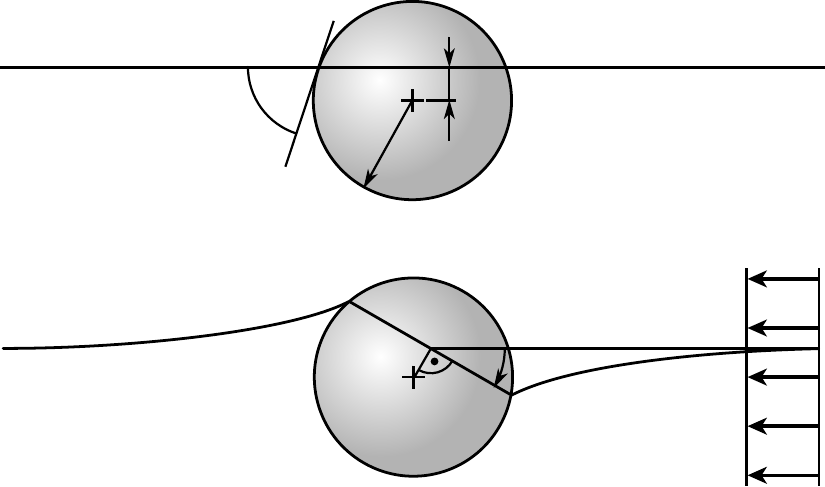}};
\coordinate (S2) at (8.5,2.26);
\coordinate (S1) at (8.5,7.95);
\path (S1) ++ (1.2,-0.6) node {$\varepsilon a$};
\path (S1) ++ (-0.4,-1.3) node {$a$};
\path (S1) ++ (-2.7,0.15) node {$\Theta$};
\path (S1) ++ (4,1.5) node(mu1) {$\mu_2$};
\path (mu1) ++ (0,-1.7) node {$\mu_1$};
\path (S1) ++ (7,0.25) node {$\Sigma_{12}$};
\path (S1) ++ (2,1.7) node {$\Sigma_{2p}$};
\path (S1) ++ (2,-1.8) node {$\Sigma_{1p}$};
\node[] at (16.2,4.8) {$U$};
\path (S2) ++ (1.45,0.3) node {$\alpha$};
\end{tikzpicture}
\caption{Sketch of a spherical particle~$p$ attached to the interface~$\Sigma_{12}$ between fluids~1 and~2: (a)~equilibrium configuration in the absence of any external force with static contact angle~$\Theta$; (b)~steady-state configuration of the driven particle moving with a constant velocity of~$U$ according to the case~$\mu_1>\mu_2$, viewed in the particle's rest frame. The particle is rotated by the trim angle~$\alpha$.}%
\label{fig:Kugel_GGW}%
\end{figure}
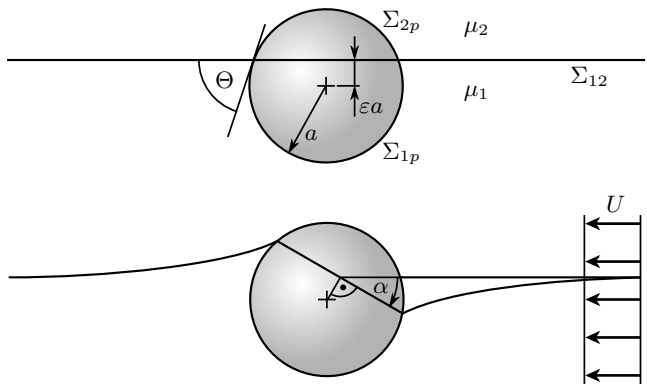
The circular three-phase contact line, TCL, is pinned to the particle surface $\Sigma_{1p}\,\cup\,\Sigma_{2p}$. Since the particle radius is the characteristic length scale of the flow field, we may neglect undulations of the TCL, which are typically tens of nanometres in size \citep{Park2011,Stamou2000}, when dealing with microparticles. For simplicity, we assume the Bond number~$\Bo\definiert\Delta\rho g a^2/\sigma_{12}$, containing the maximum density difference~$\Delta\rho$ between the three phases, the gravitational acceleration~$g$ as well as the fluid-fluid interfacial tension~$\sigma_{12}$, to be zero. The assumption~$\Bo=0$ leads to the equilibrium configuration shown in \bild{fig:Kugel_GGW}(a) characterised by a planar interface and an equilibrium contact angle~$\Theta$. Note that there is no conceptual difficulty to include a non-zero Bond number in the asymptotic model described below, since \citet{Lee1979} have demonstrated that the perturbation quantity may be a combination of the Bond number and the capillary number~$\Ca\definiert\mu_1U/\sigma_{12}$ used in the present work. In the definition of~$\Ca$, $\mu_1$ denotes the dynamic viscosity of fluid~1, while~$U$ denotes the absolute value of the undisturbed flow velocity far from the particle when the situation is analysed in the particle's rest frame (cf. \bild{fig:Kugel_GGW}(b)). The force~$\vek F$ driving the particle is supposed to act at the particle centre-of-mass and parallel to the plane given by the fluid interface in equilibrium. As discussed above, the plane containing the circular TCL has rotated by the trim angle~$\alpha$ once the particle motion has reached a steady state (see \bild{fig:Kugel_GGW}(b)). In the regime of particle Reynolds numbers~$\Rey\definiert U a/\nu_1$ much smaller than unity, where~$\nu_1$ denotes the kinematic viscosity of fluid~1, incompressible Newtonian fluid flow is governed by the Stokes equation
\begin{equation}
\vek 0=-\nabla p+\mu\nabla^2 \vek u
\label{eq:Stokes}
\end{equation}
and the continuity equation
\begin{equation}
\nabla\cdot\vek u=0,
\label{eq:Konti}
\end{equation}
where~$\vek u$ is the velocity and~$p$ is the pressure. In equations~\eqref{eq:Stokes} and~\eqref{eq:Konti}, indices~1 and~2 denoting the respective fluid phases have been omitted. On the particle surface, the flow field obeys the no-slip condition
\begin{equation}
\vek u=\vek 0~~\text{at}~\Sigma_{1p}\,\cup\,\Sigma_{2p}
\label{eq:noslip}
\end{equation}
while approaching the homogeneous velocity field~$\vek U$ far from the particle. Across the fluid-fluid interface, which is assumed pure and therefore non-viscous (zero Boussinesq number), implying absence of Marangoni stresses, the velocity is continuous
\begin{equation}
\sprung{\vek u}=\vek 0\,\Longleftrightarrow\,\vek u_1=\vek u_2~~\text{at}~\Sigma_{12}
\label{eq:vstetig}
\end{equation}
and the interfacial momentum balance
\begin{equation}
\sprung{-p\vek I+2\mu \vek S}\cdot\vek n_{12}=\sigma_{12}\left(\nabla^s\cdot\vek n_{12}\right)\vek n_{12}~~\text{at}~\Sigma_{12}
\label{eq:sprungallg}
\end{equation}
holds, where~$\vek I$ and~$\vek S$ are the unit tensor and the rate-of-strain tensor, respectively, $\vek n_{12}$ is the fluid interface normal, and $\nabla^s\definiert(\vek I-\vek n_{12}\otimes\vek n_{12})\nabla$. In equations~\eqref{eq:vstetig} and~\eqref{eq:sprungallg}, we have used the jump bracket notation
\begin{equation}
\sprung{f}(\vek r)\definiert \left[f(\vek r+\epsilon \vek n_{12})-f(\vek r-\epsilon \vek n_{12})\right]_{\epsilon\to 0}=f_1(\vek r)-f_2(\vek r)~\text{for}~\vek r\in\Sigma_{12}.
\label{eq:defsprung}
\end{equation}
Note that the last part of \gleich{eq:defsprung} implies that the orientation of the normal vector with respect to the fluid phases has been specified in such a way that it points from fluid~2 into fluid~1. Concerning the particle having the outer normal vector~$\vek n_p$, steady state is reached when the balances of linear momentum
\begin{equation}
\vek 0=\iint\limits_{\Sigma_{1p}\cup\Sigma_{2p}}\vek T\cdot\vek n_p \mathrm{d}\Sigma+\oint\limits_{\mathrm{TCL}}\sigma_{12}\left(\partiell{\vek r_\mathrm{TCL}}{s}\times\vek n_{12}\right)\mathrm{d}s+\vek F
\label{eq:impulsallg}
\end{equation}
and angular momentum with respect to the particle's centre-of-mass
\begin{equation}
\vek 0=\iint\limits_{\Sigma_{1p}\cup\Sigma_{2p}} \vek r_p\times(\vek T\cdot\vek n_p) \mathrm{d}\Sigma+\oint\limits_{\mathrm{TCL}}\vek r_\mathrm{TCL}\times\left[\sigma_{12}\left(\partiell{\vek r_\mathrm{TCL}}{s}\times\vek n_{12}\right)\right]\mathrm{d}s
\label{eq:drehmomentallg}
\end{equation}
are fulfilled. Note that the inertial contributions to the balances~\eqref{eq:impulsallg} and~\eqref{eq:drehmomentallg} have been neglected since the Reynolds number is much smaller than unity. In equations~\eqref{eq:impulsallg} and~\eqref{eq:drehmomentallg}, $\vek T\definiert-p\vek I+2\mu \vek S$ is the stress tensor, $\vek r_\mathrm{TCL}$ is the position vector of a point on the TCL, parametrized by~$s$ such that the vector~$\partial\vek r_\mathrm{TCL}/\partial s\times\vek n_{12}$ points away from the particle, and~$\vek r_p$ is the position vector of a point on the particle surface. Recall that by definition the driving force~$\vek F$ itself does not produce a torque on the particle as it acts on its centre-of-mass. We concentrate on the case of a low viscosity ratio between the two fluids, implying~$\mu_2/\mu_1\to0$, because in that case the effects studied in this work are expected to be most pronounced, and also because explicit analytical results can be found. Upon introducing the scaling
\begin{equation}
\vek u\definiertb U\vek{\tilde u}\quad p\definiertb \frac{\tilde p \mu_1 U}{a}\quad \vek S\definiertb\frac{\vek{\tilde S} U}{a}\quad\nabla\definiertb\frac{1}{a}\tilde\nabla
\label{eq:skalen}
\end{equation}
and incorporating the definition of the jump bracket~\eqref{eq:defsprung} and of the capillary number $\Ca$, we may write the interfacial momentum balance~\eqref{eq:sprungallg} as
\begin{equation}
-\tilde p_1\vek n_{12} +2 \vek{\tilde S}_1\cdot\vek n_{12}=\Ca^{-1}(\tilde\nabla^s\cdot\vek n_{12})\vek n_{12},
\label{eq:sprungmu1mu2}
\end{equation}
where the stress contribution due to fluid~2 is omitted according to the assumption of low viscosity ratio,~$\mu_2/\mu_1\to0$. The presence of the factor~$\Ca^{-1}$ on the right-hand side of \gleich{eq:sprungmu1mu2} allows for a perturbation approach in which zeroth-order stresses cause a first-order deformation of the fluid-interface, and correspondingly for higher orders. In the present work, we focus on the first-order deformation. An extension to higher orders is formally straightforward, but is not expected to yield much additional insight. For any quantity~$f$, the perturbation series for small capillary numbers is given by
\begin{equation}
f\definiertb f\ord0+f\ord1 \Ca+\landauca 2.
\label{eq:perturbf}
\end{equation}
Accordingly, the interfacial stress balance~\eqref{eq:sprungmu1mu2} becomes
\begin{equation}
\left(-\tilde p_1\ord0\vek n_{12}\ord0 +2 \vek{\tilde S}_1\ord0\cdot\vek n_{12}\ord0\right)\Ca=\left(\tilde\nabla^s\cdot\vek n_{12}\ord1\right)\vek n_{12}\ord0\Ca+\landauca2
\label{eq:sprungpert}
\end{equation}
where the planarity of the zeroth-order fluid interface, $\tilde\nabla^s\cdot\vek n_{12}\ord0=0$, has been used. 

\section{Flow field and stress distribution}

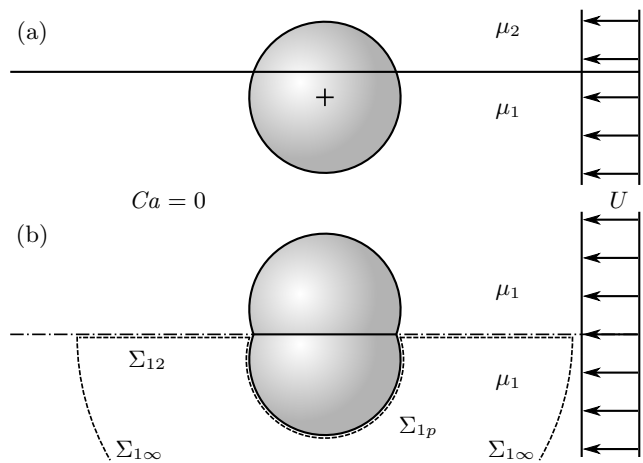
\begin{figure}
\centering%
\def\breiteb{6cm}
\begin{tikzpicture}[x=0.1*\breiteb,y=0.1*\breiteb]
\node[anchor=south west,inner sep=0pt] at (0,0) {\includegraphics[height=\breiteb]{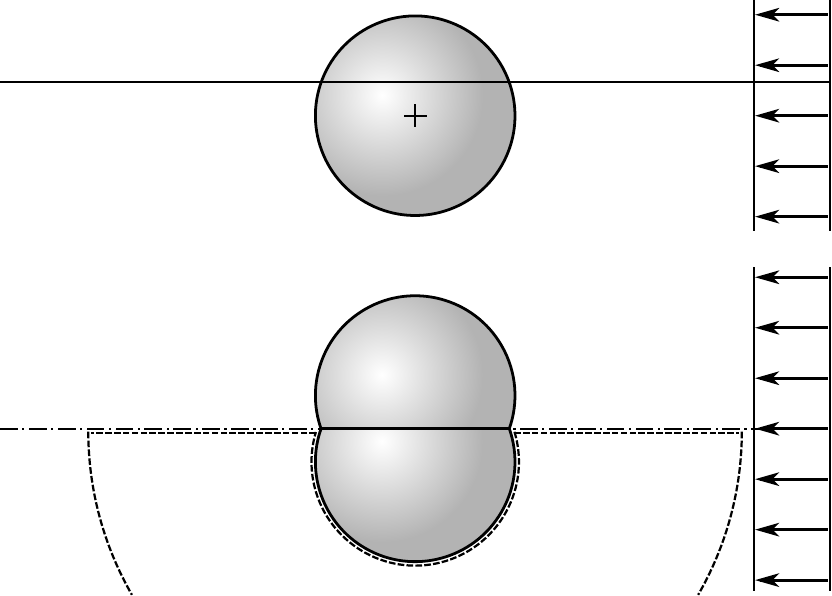}};
\coordinate (S2) at (7.05,2.75);
\coordinate (S1) at (7.05,8.05);
\path (S1) ++ (4,1.5) node(mu1) {$\mu_2$};
\path (mu1) ++ (0,-1.8) node {$\mu_1$};
\node[name=U] at (13.5,5.8) {$U$};
\path (S2) ++ (4,1) node(mu2) {$\mu_1$};
\path (mu2) ++ (0,-2) node {$\mu_1$};
\path (S2) ++ (2,-2) node {$\Sigma_{1p}$};
\path (S2) ++ (-4,-0.5) node {$\Sigma_{12}$};
\path (S2) + (4.8,-2.5) node[anchor=east] {$\Sigma_{1\infty}$} + (-4.9,-2.5) node[anchor=west] {$\Sigma_{1\infty}$};
\node at (0.5,9.5) {(a)};
\node[name=b] at (0.5,5) {(b)};
\path (U) ++ (-10,0) node {$\Ca=0$};
\end{tikzpicture}
\caption{Equivalent problems for describing the flow field within fluid~1 at zero capillary number and zero viscosity ratio: (a) original scenario including the planar fluid-fluid interface and the spherical particle; (b) transformed scenario with symmetry plane (dash-dotted) and two fused equal spheres; the control volume (dashed) utilized in the calculation of the torque is also shown.}%
\label{fig:symm}%
\end{figure}

As we are interested in the first-order curvature of the fluid interface represented by $\tilde\nabla^s\cdot\vek n_{12}\ord1$, it is obvious from \gleich{eq:sprungpert} that the flow field of order~$\Ca\ord0$ is to be described. Physically, the case~$\Ca=0$ implies that the interfacial tension is large enough to counterbalance any jump in the normal stresses across the fluid interface without noticeable interfacial deformation. Moreover, because of the low viscosity ratio assumption, the stress vector at the fluid-fluid interface is perpendicular to the latter, similar to a symmetry boundary. The flow scenario within the half-space occupied by fluid~1 at~$\Ca=0$ and~$\mu_2/\mu_1=0$ can thus be imitated by notionally removing the fluid interface and replacing fluid~2 by fluid~1. At the same time, the part of the rigid particle originally wetted by fluid~1 must be mirrored with respect to the planar fluid interface, making the latter a symmetry plane for the flow problem. Figure~\ref{fig:symm} provides a sketch of the two setups which are equivalent with respect to the flow of fluid~1.
The procedure just described allows for a convenient treatment of the flow problem based on Stokes flow around a body moving parallel to its symmetry plane (usually termed asymmetric flow as opposed to axisymmetric flow). In the case of a spherical particle with an equilibrium contact angle~$\Theta$ (cf. \bild{fig:Kugel_GGW}(a)) of~90\textdegree, the flow field of order~$\Ca^0$ is given by the well-known Stokes flow problem past a sphere \citep{Happel1983}.

\begin{figure}
\centering%
\def\breitea{5cm}
\begin{tikzpicture}[x=0.1*\breitea,y=0.1*\breitea]
\node[anchor=south west,inner sep=0pt] at (0,0) {\includegraphics[width=\breitea]{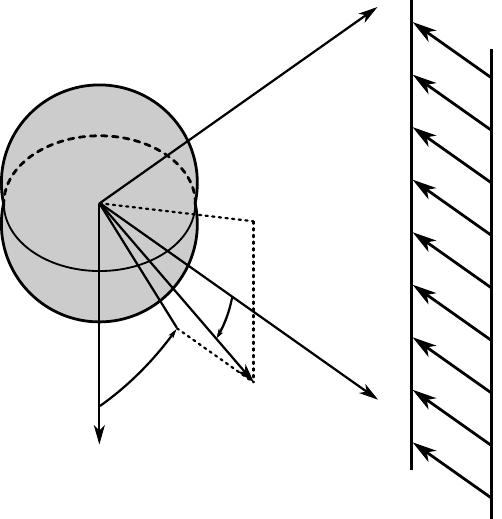}};
\node at (2.4,1.7) {$x$};
\node at (7.5,3.1) {$z$};
\node at (7.5,9.7) {$y$};
\node at (4.25,4.25) {$\theta$};
\node at (2.7,3.5) {$\varphi$};
\node at (4.6,2.6) {$\vek r$};
\node[fill=white] at (9,5.5) {$-U\boldsymbol{e}_z$};
\node[anchor=north east,text width=2.5cm] at (0,10) {$0\leq r <\infty$\\[0.5ex]$0\leq \theta <\pi$\\[0.5ex]$0\leq \varphi <2\pi$\\[0.5ex]$x=r \sin\theta\cos\varphi$\\[0.5ex] $y=r \sin\theta\sin\varphi$\\[0.5ex]$z=r\cos\theta$};
\end{tikzpicture}
\caption{Spherical Coordinates and the direction of the undisturbed flow at infinity.}%
\label{fig:Kugelkoordinaten}%
\end{figure}
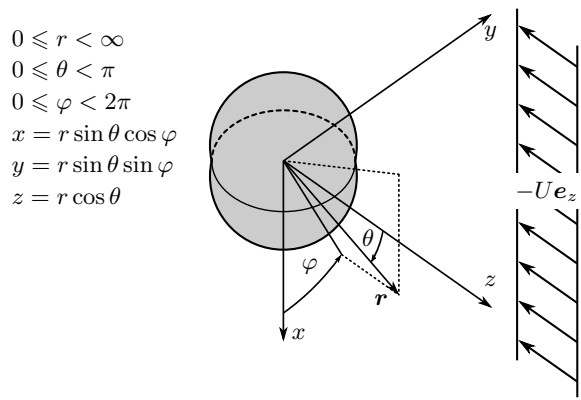

Figure~\ref{fig:Kugelkoordinaten} defines the spherical coordinate system used in the subsequent asymptotic analysis of the flow field.
Herein, the $x$-axis points downwards into the region originally occupied by fluid~1. The depicted configuration corresponds to~$\varepsilon>0$ and therefore~$\Theta<$90\textdegree, as shown in \bild{fig:Kugel_GGW}(a). Clearly, the relation
\begin{equation}
\varepsilon=\cos\Theta
\label{eq:eps}
\end{equation}
holds. The undisturbed fluid infinitely remote from the particle flows into the negative $z$-direction when~$U$ is positive, that is~$\vek U=-U\vek e_z$. A formal solution to the Stokes flow problem around two fused spheres with arbitrary~$\varepsilon$ has been developed by \citet{Zabarankin2007}. However, that solution is not suited for the present work since we intend to derive explicit analytical expressions capturing the main effects, while evaluating the results by \citet{Zabarankin2007} requires the solution of a Fredholm integral equation. Therefore, we choose an asymptotic approach developed by \citet{Brenner1964} \citep[see also][]{Happel1983}. The method is based on a representation of the particle shape (in our case, the fused spheres) as a power series in a geometric perturbation quantity (in our case, $\varepsilon$) by means of surface spherical harmonics. In this way, the no-slip boundary condition~\eqref{eq:noslip} on the generally non-spherical particle surface may be approximated by a more complicated velocity boundary condition on a spherical particle surface. Upon fitting the general solution of the Stokes equation~\eqref{eq:Stokes}, as given by \citet{Lamb1932}, in terms of \emph{solid} spherical harmonics, to the boundary condition projected onto the spherical surface written in terms of \emph{surface} spherical harmonics, the specific solution to the flow problem can be determined. Such an asymptotic approach implies that we focus on a situation in which the contact angle~$\Theta$ is not too far from 90\textdegree{}.

Consulting \bild{fig:Kugelkoordinaten} and \ref{fig:Kugel_GGW}(a), we see that the coordinates of a point~$\vek r_p= r_p\vek e_r=x_p \vek e_x +y_p\vek e_y +z_p\vek e_z$ on the particle-fluid interface obey the relation
\begin{equation}
(x_p-\varepsilon a)^2+y_p^2+z_p^2=a^2
\label{eq:rpkart}
\end{equation}
in the Cartesian system, and consequently
\begin{equation}
\tilde r_p=\varepsilon \sin\theta\left|\cos\varphi\right|+\sqrt{\varepsilon^2 \left(\sin^2\theta\cos^2\varphi-1\right)+1}\sim 1+\varepsilon \sin\theta\left|\cos\varphi\right|
\label{eq:rpkugel1}
\end{equation}
in spherical coordinates and non-dimensional quantities. Within the scope of the present work, it is sufficient to restrict the analysis to first order in~$\varepsilon$. Additionally, in order to simplify notation, we usually omit higher-order terms in~$\varepsilon$, keeping in mind that the subsequent expressions containing~$\varepsilon$ are valid only up to first order. The particle shape is therefore approximately given by
\begin{equation}
\tilde r_p=1+\varepsilon \sin\theta\left|\cos\varphi\right|.
\label{eq:rplin}
\end{equation}
According to the method by \citet{Brenner1964}, we need to expand the last term  in \gleich{eq:rplin} into a series of surface spherical harmonics,
\begin{equation}
\sin\theta\left|\cos\varphi\right|= \sum_{k=0}^{\infty}f_k(\theta,\varphi),
\label{eq:entwssh1}
\end{equation}
where~$f_k$ is a surface spherical harmonic of order~$k$ defined by
\begin{equation}
\nabla^2\left(r^k f_k\right)=0.
\label{eq:fkdef}
\end{equation}
In more detail, using the associated Legendre polynomials~$P_k^m(\cos\theta)$, we write \citep{Byerly1983,Dassios2001}
\begin{equation}
f_k(\theta,\varphi)=\sum_{m=0}^{k}c_{km}N_{km}P_k^m(\cos\theta)\cos(m\varphi)
\label{eq:fkdetail}
\end{equation}
with the normalization coefficients
\begin{equation}
N_{km}=
\begin{cases}
\sqrt{2} \sqrt{\frac{2k+1}{4\pi}\frac{(k-m)!}{(k+m)!}} &~\text{if}~m>0\\
\sqrt{\frac{2k+1}{4\pi}} &~\text{if}~m=0.\\
\end{cases}
\label{eq:Nkm}
\end{equation}
The expansion coefficients~$c_{km}$ are given by
\begin{equation}
c_{km}=\int \limits_{0}^{\pi} \int \limits_{0}^{2\pi} \sin\theta\left|\cos\varphi\right| N_{km} P_k^m(\cos\theta)\cos(m\varphi)\sin\theta\mathrm{d}\theta\mathrm{d}\varphi.
\label{eq:ckm}
\end{equation}
We have omitted the terms corresponding to~$m<0$, associated with~$\sin(m\varphi)$, in equations~\eqref{eq:fkdetail} and~\eqref{eq:Nkm} since the respective~$c_{km}$ can be shown to vanish in the expansion~\eqref{eq:entwssh1}. In table~\ref{tab:ckm}, the first few expansion coefficients are listed.
\begin{table}%
\renewcommand{\arraystretch}{1.5}
\centering%
\begin{tabular}{lclclll}
\toprule
 &  & \multicolumn{1}{c}{$c_{km}$} & \hphantom{xxxx}& & & \multicolumn{1}{c}{$c_{km}$}\\
\toprule
$k=0$ & $m = 0$\hphantom{0} & $\sqrt{\pi}$& & $k=8$ & $m=0$ & $-35\sqrt{17\pi}/16384$ \\
$k=2$ & $m=0$ & $-\sqrt{5\pi}/8$& & & $m=2$ & $3\sqrt{595\pi/2}/4096$ \\
 &  $m=2$ &  $\sqrt{15\pi}/8$& & & $m=4$ & $-3\sqrt{1309\pi}/8192$\\
 $k=4$ & $m=0$ & $-3\sqrt{\pi}/64$& & & $m=6$ & $\sqrt{7293\pi/2}/4096$  \\
 & $m=2$ &  $\sqrt{5\pi}/32$& & & $m=8$ & $-3\sqrt{12155\pi}/16384$\\
 &  $m=4$ & $-\sqrt{35\pi}/64$& & $k=10$ & $m=0$ & $-147\sqrt{21\pi}/131072$ \\
 $k=6$& $m=0$ & $ -5\sqrt{13\pi}/1024$& & & $m=2$ & $49\sqrt{385\pi}/131072$ \\
 & $m=2$ & $\sqrt{1365\pi/2}/1024$& & & $m=4$ & $-7\sqrt{5005\pi}/65536$ \\
 & $m=4$ &  $-3\sqrt{91\pi}/1024$& & & $m=6$ & $21\sqrt{5005\pi/2}/131072$\\
 & $m=6$ &  $\sqrt{3003\pi/2}/1024$& & & $m=8$ & $-7\sqrt{85085\pi/3}/131072$\\
 &  &  & & & $m=10$ & $7\sqrt{323323\pi/6}/131072$\\
\bottomrule
\end{tabular}
\caption{Expansion coefficients~$c_{km}$ for two fused spheres approximated to first order in $\varepsilon$ as calculated from \gleich{eq:ckm}; coefficients with $m<0$ and/or $m$ odd vanish.}
\label{tab:ckm}
\end{table}
Now that we know the spherical harmonics expansion of the~$\landaueps{}$-approximation of two fused spheres, we turn to the calculation of the solution
\begin{equation}
\vek{\tilde{u}}\ord{0}\definiertb\vek{\tilde{u}}\ord{0,0} +\varepsilon\vek{\tilde{u}}\ord{0,1}+\landaueps2
\label{eq:uentw}
\end{equation}
and analogously for the pressure~$\tilde p\ord{0}$. Recalling that~$\varepsilon=0$ describes the case of a particle with a contact angle of 90\textdegree{} (or, equivalently, two fused spheres collapsed into a single sphere), it is obvious that
\begin{align}
\vek{\tilde{u}}\ord{0,0}&=-\frac{1}{2}\cos\theta\left(\frac{1}{\tilde r^3}-\frac{3}{\tilde r}+2\right)\vek e_r-\frac{1}{4}\sin\theta\left(\frac{1}{\tilde r^3}+\frac{3}{\tilde r}-4\right)\vek e_\theta\label{eq:u00}\\
\tilde{p}\ord{0,0}&=\frac{3}{2}\frac{\cos\theta}{\tilde r^2} \label{eq:p00}
\end{align}
which is the solution to the classical problem of Stokes flow past a sphere \citep{Happel1983}. Regarding the $\landaueps{}$-term in \gleich{eq:uentw}, \citet{Brenner1964} has constructed the boundary condition
\begin{equation}
\left.\vek{\tilde{u}}\ord{0,1}\right|_{\tilde r =1}=-\sin\theta|\cos\phi|\left.\partiell{\vek{\tilde{u}}\ord{0,0}}{\tilde r}\right|_{\tilde r=1}=-\sum_{k=0}^{\infty}f_k(\theta,\varphi)\left.\partiell{\vek{\tilde{u}}\ord{0,0}}{\tilde r}\right|_{\tilde r=1}
\label{eq:rbv01}
\end{equation}
projected onto a single sphere. The formalism leading to the velocity and pressure fields~$\vek{\tilde{u}}\ord{0,1}$ and~$\tilde p\ord{0,0}$ can be found elsewhere \citep{Brenner1964,Happel1983}. But, although that method can in principle be directly applied to the problem of fused spheres, corrections to certain formulae given in the above references are needed. If the formalism is applied consistently with \citet{Brenner1964} and \citet{Happel1983}, the results for the velocity and pressure fields do not solve the Stokes equation. For the special case~$\vek U=-U\vek e_z$ discussed here, we will thus provide the corrected expressions using the same notation as \citet{Brenner1964}. To be precise, equations~(3.22) and~(3.23) given by \citet{Brenner1964} need to be reformulated\label{page:Brenner}. The goal is to express the right-hand side of
\begin{equation}
\sum_{n=1}^{\infty}\kYn=\frac{3}{2}\left(\tilde r \vek{\tilde{U}}\cdot\tilde\nabla f_k-2f_k \vek{\tilde{U}}\cdot\vek e_r\right)
\label{eq:318}
\end{equation} 
through a sum of surface spherical harmonics to find the~$\kYn$. Since~$\vek{\tilde{U}}=-\vek e_z$ in the present work, after some calculation we arrive at
\begin{equation}
\begin{split}
\sum_{n=1}^{\infty}\kYn=-\frac{3}{2}\sum_{m=0}^{k}\frac{c_{km}N_{km}\cos(m\varphi)}{2k+1}&\left[(1-k)(k+m)P_{k-1}^m(\cos\theta)\right. \\&\left.+(2+k)(k-m+1)P_{k+1}^m(\cos\theta)\right]
\label{eq:318richtig}
\end{split}
\end{equation}
replacing equation~(3.22) by \citet{Brenner1964} and allowing us to read off
\begin{equation}
\kYn=
\begin{cases}
\dfrac{3}{2}\displaystyle\sum_{m=0}^{k}\dfrac{c_{km}N_{km}\cos(m\varphi)}{2k+1}(k-1)(k+m)P_{k-1}^m(\cos\theta) & \text{for}~n=k-1\\
-\dfrac{3}{2}\displaystyle\sum_{m=0}^{k}\dfrac{c_{km}N_{km}\cos(m\varphi)}{2k+1}(2+k)(k-m+1)P_{k+1}^m(\cos\theta) & \text{for}~n=k+1\\
0 & \text{for all other}~n.\\
\end{cases}
\label{eq:kYn}
\end{equation}
In conjunction with equations~(2.12a-c) by \citet{Brenner1964} and Lamb's solution \citep[equations~(2.13a,b)]{Brenner1964}, \gleich{eq:kYn} yields the expressions for~$\vek{\tilde{u}}\ord{0,1}$ and~$\tilde p\ord{0,1}$. As those expressions are very lengthy, we waive writing them down at this point and go over to deriving the force and torque acting on the particle as well as the stress at the fluid-fluid interface.
Due to the symmetry, the drag force~$\vek F_D\ord0$ on the particle in \bild{fig:symm}(a) is simply half the drag force on the mirrored particle in \bild{fig:symm}(b). The latter can be calculated based on equation~(2.20) by \citet{Brenner1964}, which is well-suited for the perturbation approach used above, and is found to be
\begin{equation}
\vek F_D\ord0=-3\pi\mu_1 U a\left(1+\frac{9}{16}\varepsilon\right)\vek e_z.
\label{eq:FD}
\end{equation}
As expected, if~$\varepsilon=0$, the drag is half the Stokes drag of a sphere \citep{Radoev1992,Danov1995,Ally2010}. Equally intuitive is the increase in drag with increasing~$\varepsilon$, that is deeper immersion of the particle in fluid~1. We now turn to the torque acting on the particle, occurring in the balance of angular momentum~\eqref{eq:drehmomentallg} and therefore required to determine the trim angle~$\alpha$. Instead of directly calculating the torque on the particle, we write down the $y$-component of the balance of angular momentum for the control volume enclosed by the interfaces~$\Sigma_{12},~\Sigma_{1p}$ and~$\Sigma_{1\infty}$ shown in \bild{fig:symm}(b), which reads
\begin{equation}
0=\vek e_y\cdot\left[\iint\limits_{\Sigma_{12}} \vek r\times(\vek T\ord0\cdot\vek n\ord0) \mathrm{d}\Sigma+\iint\limits_{\Sigma_{1p}} \vek r\times(\vek T\ord0\cdot\vek n\ord0) \mathrm{d}\Sigma+\iint\limits_{\Sigma_{1\infty}} \vek r\times(\vek T\ord0\cdot\vek n\ord0) \mathrm{d}\Sigma\right],
\label{eq:drehmomentfluid}
\end{equation}
where the vectors~$\vek n\ord0$ are outer normals with respect to the control volume. Again, inertial terms have been neglected according to the assumption of small Reynolds numbers. The second integral on the right-hand side of \gleich{eq:drehmomentfluid} is equal to the negative of the viscous torque acting on the particle (from \gleich{eq:drehmomentallg}) in the case~$\mu_2/\mu_1=0$. The third integral can be shown to vanish as the radius of the hemispherical interface~$\Sigma_{1\infty}$ goes to infinity. We therefore only need to calculate the integral over the fluid-fluid interface~$\Sigma_{12}$ whose integrand contains the stress distribution on~$\Sigma_{12}$ (which is equal to the stress difference across the interface when~$\mu_2/\mu_1=0$). Conveniently, the same stress distribution is required for the calculation of the $\landauca{}$ interfacial deformation further below. We have already noted that the fluid-fluid interface of order~$\Ca^0$ is a symmetry boundary to which the stress vector is perpendicular. Since for~$\Ca=0$ we see from \bild{fig:Kugelkoordinaten} that the outer normal at~$\Sigma_{12}$ is~$\vek e_\varphi$ at~$\varphi=\pi/2$ and~$-\vek e_\varphi$ at~$\varphi=3\pi/2$, only the component~$T_{\varphi\varphi}\ord0$ is non-zero. The latter is given by
\begin{equation}
\label{eq:Tff}
\begin{split}
\tilde T_{\varphi\varphi}\ord0= \sum_{\nu=1}^{\infty}\frac{\tau_\nu\ord0}{\tilde r^{2\nu+2}}\cos\theta= & \left[-\left(\frac{3}{2}+\frac{33}{8}\varepsilon\right)\frac{1}{\tilde r^4}+\frac{1485}{512}\frac{\varepsilon}{\tilde r^6}+\frac{105}{128}\frac{\varepsilon}{\tilde r^8}+\frac{28035}{65536}\frac{\varepsilon}{\tilde r^{10}}\right.\\%
& +\frac{35343}{131072}\frac{\varepsilon}{\tilde r^{12}}+\frac{3144141}{1677216}\frac{\varepsilon}{\tilde r^{14}}+\frac{2323035}{16777216}\frac{\varepsilon}{\tilde r^{16}}\\
&\left.+\frac{485474445}{4294967296}\frac{\varepsilon}{\tilde r^{18}}+\frac{729555255}{8589934592}\frac{\varepsilon}{\tilde r^{20}}+\dotsb\right]\cos\theta \\
\end{split}
\end{equation}
at both~$\varphi=\pi/2$ and~$\varphi=3\pi/2$. Remarkably, the stress distribution has the form of an infinite series in negative powers of~$\tilde r$. As a consequence, we need to check the convergence of any quantity dependent on~$\tilde T_{\varphi\varphi}\ord0$. Now we are in a position to evaluate
\begin{align}
\vek e_y\cdot\iint\limits_{\Sigma_{12}}\vek r&\times(\vek T\ord0\cdot\vek n\ord0) \mathrm{d}\Sigma =2\int\limits_{a\sqrt{1-\varepsilon^2}}^{\infty}\int\limits_{0}^{\pi}\vek e_y
\cdot\left[r\vek e_r\times\left(\vek T\ord0\cdot\vek e_\varphi\right)\right]_{\varphi=\pi/2}r\mathrm{d}\theta\mathrm{d}r\\
  &=2\int\limits_{a}^{\infty}\int\limits_{0}^{\pi}\vek e_y
\cdot\left[r\vek e_r\times\left(\vek T\ord0\cdot\vek e_\varphi\right)\right]_{\varphi=\pi/2}r\mathrm{d}\theta\mathrm{d}r+\landaueps{2} \\
  &=2\int\limits_{a}^{\infty}\int\limits_{0}^{\pi}\vek e_y
\cdot\left[r\vek e_r\times\left(T_{\varphi\varphi}\ord0\vek e_\varphi\right)\right]_{\varphi=\pi/2}r\mathrm{d}\theta\mathrm{d}r+\landaueps{2} \\
 &=-2\mu_1 U a^2\int\limits_{1}^{\infty}\int\limits_{0}^{\pi}\left.\tilde{T}_{\varphi\varphi}\ord0\right|_{\varphi=\pi/2}\tilde r^2\cos\theta\mathrm{d}\theta\mathrm{d}\tilde r\quad(\text{to order}~\varepsilon)\\
&\overset{\mathclap{\eqref{eq:Tff}}}{=} \hphantom{+}\pi\mu_1 U a^2\sum_{\nu=1}^{\infty}\frac{\tau_\nu\ord0}{2\nu-1}\label{eq:viskmoment}
\end{align}
which will be used further below for finding the trim angle~$\alpha$.

\section{Interfacial deformation and particle trim angle}

\begin{figure}
\centering%
\begin{tikzpicture}[every axis/.append style={
	xmin=-3,
	xmax=3,
	ymin=-0.3,
	ymax=0.3,
	xtick={-3,-2,-1,0,1,2,3},
	ytick=\empty,
	extra y ticks={-0.3,0,0.3},
	extra y tick labels={0.3,0,-0.3},
	x=0.12\textwidth,
	y=0.12\textwidth,
	xlabel={$\tilde z$},
	ylabel={$\tilde x$},
	}]
\begin{axis}[title={$\Theta=75$\textdegree\vphantom{\Huge A}}
	]
\addplot[gray,densely dashed] coordinates {(-1,-1) (-1,1)};
\addplot[gray,densely dashed] coordinates {(1,-1) (1,1)};
\addplot[color=blue] file {Schnitt_Theta75_daten_1.dat};
\addplot[color=blue] file {Schnitt_Theta75_daten_2.dat};
\addplot[color=blue] file {Schnitt_Theta75_daten_3.dat};
\end{axis}
\begin{axis}[title={$\Theta=90$\textdegree},
yshift=-0.2\textwidth,
	]
\addplot[gray,densely dashed] coordinates {(-1,-1) (-1,1)};
\addplot[gray,densely dashed] coordinates {(1,-1) (1,1)};
\addplot[color=blue] file {Schnitt_Theta90_daten_1.dat};
\addplot[color=blue] file {Schnitt_Theta90_daten_2.dat};
\addplot[color=blue] file {Schnitt_Theta90_daten_3.dat};
\end{axis}
\begin{axis}[title={$\Theta=105$\textdegree},
yshift=-0.4\textwidth,
]
\addplot[gray,densely dashed] coordinates {(-1,-1) (-1,1)};
\addplot[gray,densely dashed] coordinates {(1,-1) (1,1)};
\addplot[color=blue] file {Schnitt_Theta105_daten_1.dat};
\addplot[color=blue] file {Schnitt_Theta105_daten_2.dat};
\addplot[color=blue] file {Schnitt_Theta105_daten_3.dat};
\end{axis}
\end{tikzpicture}%
\caption{Cross section of the interfacial shape~$\tilde h=\tilde h\ord1\Ca$ according to equation~(\ref{eq:hlsg2}) and of the corresponding three-phase contact line for three different equilibrium contact angles~$\Theta$ and a capillary number of~$\Ca=0.3$; the inclination of the TCL is given by the particle trim angle~$\alpha=\alpha\ord1\Ca$ from equation~(\ref{eq:alpha1}); the particle is moving in the~$\tilde z$-direction.}%
\label{fig:Schnitt}%
\end{figure}
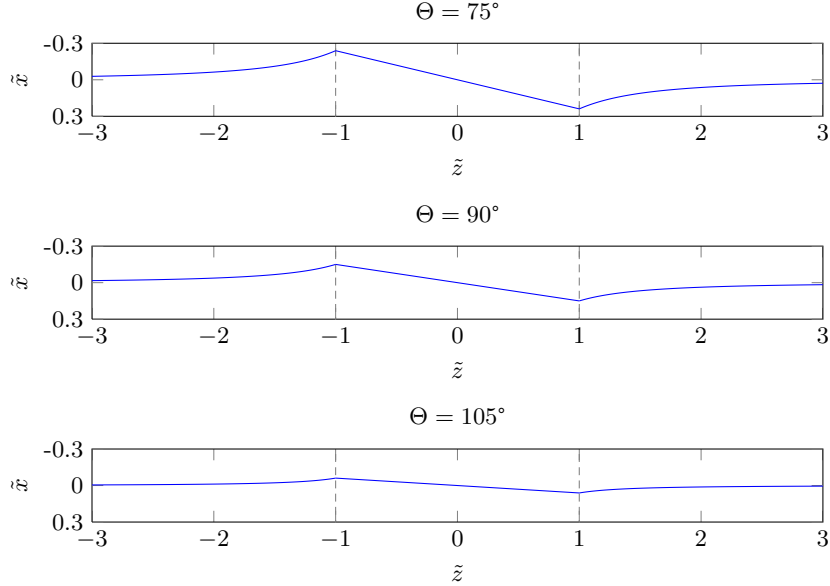

In order to parametrize the shape of the fluid-fluid interface~$\Sigma_{12}$ of order~$\Ca$ -- resulting, within the perturbation approach, from the stress distribution of order~$\Ca^0$ calculated above -- we introduce circular cylindrical coordinates~$(\varrho,\vartheta,\chi)$
\begin{equation}
\varrho=\left.r\right|_{x=0},~\vartheta=2\pi-\theta,~\text{and}~\chi=x=r\sin\theta\cos\varphi,
\label{eq:zyl}
\end{equation}
allowing us to utilize the expression~\eqref{eq:Tff} for the stress distribution by setting~$\tilde r\to\tilde \varrho$ and~$\theta\to2\pi-\vartheta$. Note that in the cylindrical coordinate system~$0\leq\vartheta< 2\pi$, in contrast to~$0\leq\theta<\pi$ for spherical coordinates, which does not introduce a sign error but helps us to avoid the distinction between the regions of positive and negative~$y$ (that is, between~$\varphi=\pi/2$ and~$\varphi=3\pi/2$ at~$x=0$) necessary in the context of spherical coordinates (cf. \gleich{eq:Spannungzyl} below). Within the cylindrical coordinate system, we measure the interfacial deformation through its undulation in the direction of the~$\chi$-axis,
\begin{equation}
\vek r_{12}=\varrho\vek e_\varrho+h(\varrho,\vartheta)\vek e_\chi.
\label{eq:para}
\end{equation}
By choosing the orientation of the normal vector~$\vek n_{12}$ consistently with the jump condition~\eqref{eq:sprungallg}, we obtain
\begin{equation}
\vek n_{12}=\frac{\partiell{\vek r_{12}}{\varrho}\times\partiell{\vek r_{12}}{\vartheta}}{\left\|\partiell{\vek r_{12}}{\varrho}\times\partiell{\vek r_{12}}{\vartheta}\right\|}=\frac{-\partiell{h}{\varrho}\vek e_\varrho-\frac{1}{\varrho}\partiell{h}{\vartheta}\vek e_\vartheta+\vek e_\chi}{\sqrt{1+\left(\partiell{h}{\varrho}\right)^2+\left(\frac{1}{\varrho}\partiell{h}{\vartheta}\right)^2}}
\label{eq:n12}
\end{equation}
and insert the expansion~$h=h\ord1\Ca+\landauca2$. We are lead to
\begin{equation}
\vek n_{12}=\vek n_{12}\ord0+\vek n_{12}\ord1\Ca+\landauca2=\vek e_\chi-\left(\partiell{h\ord1}{\varrho}\vek e_\varrho+\frac{1}{\varrho}\partiell{h\ord1}{\vartheta}\vek e_\vartheta\right)\Ca+\landauca2
\label{eq:n12lin}
\end{equation}
as well as to the curvature
\begin{equation}
\nabla^s\cdot\vek n_{12}=-\left(\frac{\partial^2 h\ord1}{\partial\varrho^2}+\frac{1}{\varrho}\partiell{h\ord1}{\varrho}+\frac{1}{\varrho^2}\frac{\partial^2h\ord1}{\partial\vartheta^2}\right)\Ca+\landauca2.
\label{eq:divsn12}
\end{equation}
In order to evaluate the interfacial momentum balance~\eqref{eq:sprungpert}, we need to specify the sign of the stress deforming the interface. Since the left-hand side of \gleich{eq:sprungpert} is equal to the term~$\vek{\tilde{T}}\ord0\cdot\vek n_{12}\ord0$ and we are interested in the normal component of the stress, we may write (using~$\vek n_{12}\ord0=\vek e_\chi=\vek e_x$ according to \gleich{eq:n12lin})
\begin{equation}
\left(\vek{\tilde{T}}\ord0\cdot\vek n_{12}\ord0\right)\cdot\vek n_{12}\ord0=
\begin{cases}
\left(-\vek{\tilde{T}}\ord0\cdot\vek e_\varphi\right)\cdot(-\vek e_\varphi)&=\tilde T_{\varphi\varphi}\ord0~\text{for } \varphi=\pi/2 \\[2ex]
\left(\vek{\tilde{T}}\ord0\cdot\vek e_\varphi\right)\cdot\vek e_\varphi&=\tilde T_{\varphi\varphi}\ord0~\text{for } \varphi=3\pi/2. 
\end{cases}
\label{eq:Spannungzyl}
\end{equation}
Insertion of equations~\eqref{eq:divsn12}, \eqref{eq:Spannungzyl}, and~\eqref{eq:Tff} -- now in cylindrical coordinates -- into the jump condition~\eqref{eq:sprungpert} and scalar multiplication with~$\vek n_{12}\ord0$ yields
\begin{equation}
-\sum_{\nu=1}^{\infty}\frac{\tau_\nu\ord0}{\tilde \varrho^{2\nu+2}}\cos\vartheta=\frac{\partial^2 \tilde h\ord1}{\partial\tilde\varrho^2}+\frac{1}{\tilde\varrho}\partiell{\tilde h\ord1}{\tilde\varrho}+\frac{1}{\tilde\varrho^2}\frac{\partial^2\tilde h\ord1}{\partial\vartheta^2}.
\label{eq:sprungend}
\end{equation}
In addition, boundary conditions for~$h$ are to be specified. At~$\varrho\to\infty$, obviously we have~$h\to0$ because the origin of the cylindrical coordinate system coincides with the centre of the disk enclosed by the TCL (see \bild{fig:Ellipse}). A closer look is needed concerning the boundary condition on the TCL which has an elliptic shape with principal axes~$2a\sqrt{1-\varepsilon^2}$ and~$2a\sqrt{1-\varepsilon^2}\cos\alpha$ when projected onto the~$yz$-plane.
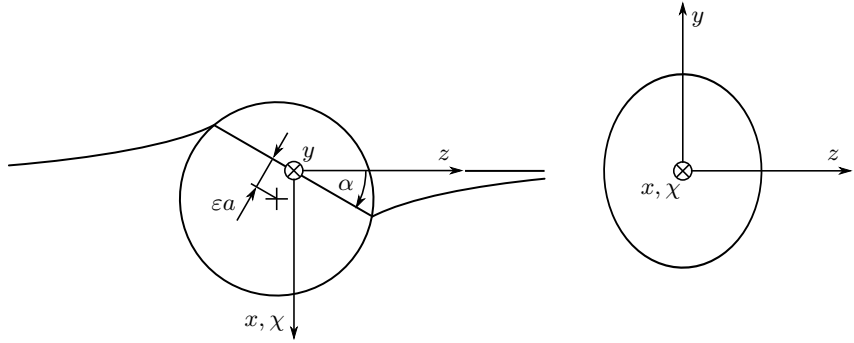
\begin{figure}
\centering%
\def\breitea{4.5cm}
\begin{tikzpicture}[x=0.1*\breitea,y=0.1*\breitea]
\node[anchor=south west,inner sep=0pt] at (0,0) {\includegraphics[height=\breitea]{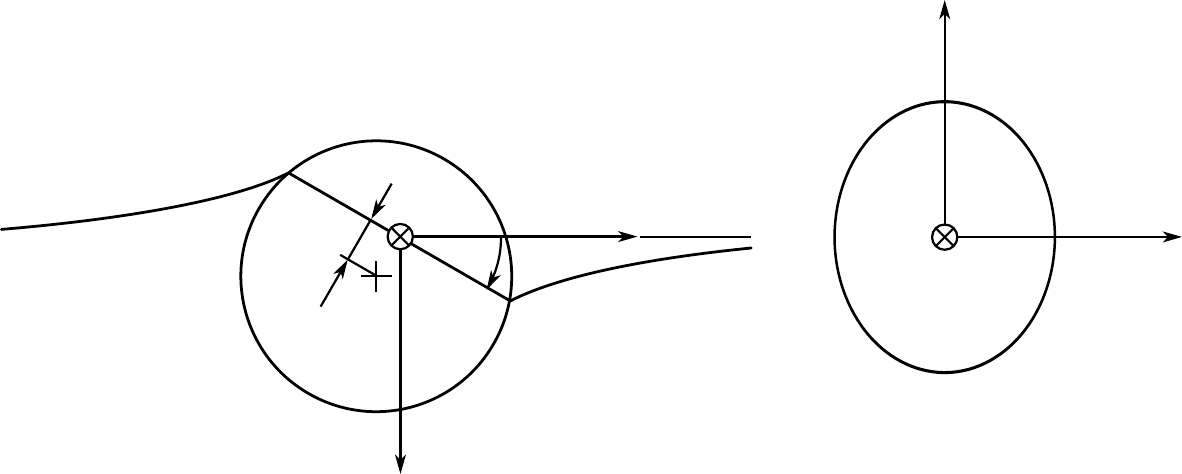}};
\coordinate (S1) at (8.4,5);
\path (S1) ++ (1.6,-0.4) node {$\alpha$};
\path (S1) ++ (4.5,0.5) node {$z$};
\path (S1) ++ (-0.8,-4.5) node {$x,\chi$};
\path (S1) ++ (0.5,0.5) node {$y$};
\path (S1) ++ (-1.4,-1.) node[anchor=east] {$\varepsilon a$};
\coordinate (S2) at (19.9,5);
\path (S2) ++ (4.5,0.5) node {$z$};
\path (S2) ++ (0.5,4.5) node {$y$};
\path (S2) ++ (-0.6,-0.6) node {$x,\chi$};
\end{tikzpicture}
\caption{Geometry of the three-phase contact line (TCL) when both~$\alpha$ and~$\varepsilon$ are positive.}%
\label{fig:Ellipse}%
\end{figure}
The coordinates of a point~$\vek r_\mathrm{TCL}$ on the TCL can be found by first describing a circle of radius~$a\sqrt{1-\varepsilon^2}$ embedded in the $yz$-plane and, secondly, by rotating the coordinate system by an angle of~$\alpha$ around the~$y$-axis. One obtains
\begin{equation}
\vek r_\mathrm{TCL}=a\sqrt{\frac{1-\varepsilon^2}{\sin^2\vartheta\cos^2\alpha+\cos^2\vartheta}}\left(\cos\alpha\vek e_\varrho+\cos\vartheta\sin\alpha\vek e_\chi\right).
\label{eq:rtcl}
\end{equation}
Since the trim angle~$\alpha$ depends on the particle velocity, it can be written as~$\alpha=\alpha\ord1\Ca+\landauca2$. In conjunction with~$\sqrt{1-\varepsilon^2}=1+\landaueps2$ and by omission of the terms~$\landaueps2$, \gleich{eq:rtcl} is simplified to
\begin{equation}
\vek{\tilde{r}}_\mathrm{TCL}=\vek e_\varrho+\alpha\ord1\cos\vartheta\Ca\vek e_\chi+\landauca2.
\label{eq:rtcllin}
\end{equation}
Considering the parametrization~\eqref{eq:para} and expanding
\begin{equation}
\left.\tilde h\ord1\right|_{\mathrm{TCL}}=\tilde h\ord1(1,\vartheta)+\landauca{},
\label{eq:hentwTCL}
\end{equation}
we read off the boundary condition
\begin{equation}
\tilde h\ord1(1,\vartheta)=\alpha\ord1\cos\vartheta.
\label{eq:htcl}
\end{equation}
It is now straightforward to find the solution to \gleich{eq:sprungend} consistent with the boundary conditions~\eqref{eq:htcl} and~$\tilde h\ord1(\tilde\varrho\to\infty,\vartheta)=0$, which is given by
\begin{equation}
\tilde h\ord1(\tilde\varrho,\vartheta)=\left[\frac{\alpha\ord1}{\tilde\varrho}+\sum_{\nu=1}^{\infty}\frac{\tau_\nu\ord0}{1-4\nu^2}\left(\frac{1}{\tilde\varrho^{2\nu}}-\frac{1}{\tilde\varrho}\right)\right]\cos\vartheta.
\label{eq:hlsg}
\end{equation}
Note that the trim angle~$\alpha\ord1$ is still unknown and is to be determined from the angular momentum balance~\eqref{eq:drehmomentallg}. Owing to the identical stress term~$\partial\vek r_\mathrm{TCL}/\partial s\times\vek n_{12}$ in equations \gleich{eq:impulsallg} and~\eqref{eq:drehmomentallg} we can derive the expression for~$\vek F$ in parallel to the one for~$\alpha\ord1$. The parametrization is chosen as~$\mathrm{d}\tilde s=\mathrm{d}\vartheta$ (recall that the stress vector shall be oriented outwards with respect to the particle). Since~$\partial\vek e_\varrho/\partial\vartheta=\vek e_\vartheta$, we may deduce from \gleich{eq:rtcllin}
\begin{equation}
\partiell{\vektilde r_\mathrm{TCL}}{\tilde s}=\partiell{\vek{\tilde{r}}_\mathrm{TCL}}{\vartheta}=\vek e_\vartheta-\alpha\ord1\sin\vartheta\vek e_\chi\Ca+\landauca2.
\label{eq:drtcl}
\end{equation}
The fluid-fluid interfacial normal vector~$\vek n_{12}$, or -- as a consequence of \gleich{eq:n12lin} -- the derivatives of~$\tilde h\ord1$ with respect to~$\tilde\varrho$ and~$\vartheta$ need to be evaluated at the TCL. Analogously to \gleich{eq:hentwTCL}, this means that the derivatives of~$\tilde h\ord1$ in \gleich{eq:n12lin} are to be evaluated at~$(1,\vartheta)$ as long as we limit the analysis to first order in~$\Ca$. As a result, we find
\begin{equation}
\partiell{\vektilde r_\mathrm{TCL}}{\tilde s}\times\vek n_{12}=\vek e_\varrho+\Ca \left.\partiell{\tilde h\ord1}{\tilde\varrho}\right|_{(1,\vartheta)}\overset{\eqref{eq:hlsg}}{=}\vek e_\varrho-\Ca\left(\alpha\ord1-\sum_{\nu=1}^{\infty}\frac{\tau_\nu\ord0}{1+2\nu}\right)\cos\theta\vek e_\chi.
\label{eq:sigmaspannung}
\end{equation}
Obviously, the integral along the TCL occurring in \gleich{eq:impulsallg} vanishes, which implies that interfacial tension does not contribute to the drag. The driving force~$\vek F\ord0$ is therefore equal to the negative of the drag force~$\vek F_D\ord0$. In the case of the~$y$-component of \gleich{eq:drehmomentallg}, normalized by~$\sigma_{12}a^2$, a similar calculation using the result~\eqref{eq:sigmaspannung} yields
\begin{equation}
\vek e_y\cdot\oint\limits_{\mathrm{TCL}}\vektilde r_\mathrm{TCL}\times\left(\partiell{\vektilde r_\mathrm{TCL}}{\tilde s}\times\vek n_{12}\right)\mathrm{d}\tilde s=-\pi\Ca\left(2\alpha\ord1-\sum_{\nu=1}^{\infty}\frac{\tau_\nu\ord0}{1+2\nu}\right).
\label{eq:sigmamoment}
\end{equation}
As has been discussed above in the context of \gleich{eq:drehmomentfluid}, the viscous torque acting on the particle is equal to the negative of the viscous torque~\eqref{eq:viskmoment} acting on the fluid volume enclosed by the interfaces~$\Sigma_{12},~\Sigma_{1p}$ and~$\Sigma_{1\infty}$ (recall \bild{fig:symm}(b)):
\begin{equation}
\vek e_y\cdot\iint\limits_{\Sigma_{1p}\cup\Sigma_{2p}} \vektilde r_p\times(\vektilde T\ord0\cdot\vek n_p) \mathrm{d}\tilde\Sigma=-\vek e_y\cdot\iint\limits_{\Sigma_{12}}\vektilde r\times(\vektilde T\ord0\cdot\vek n\ord0) \mathrm{d}\tilde\Sigma =\pi\Ca\sum_{\nu=1}^{\infty}\frac{\tau_\nu\ord0}{1-2\nu}.
\label{eq:pviskmoment}
\end{equation}
The $y$-component of \gleich{eq:drehmomentallg} can now be solved by insertion of the expressions~\eqref{eq:sigmamoment} and~\eqref{eq:pviskmoment}. We are led to the result for the particle trim angle to first order in~$\Ca$
\begin{equation}
\alpha=\sum_{\nu=1}^{\infty}\frac{\tau_\nu\ord0}{1-4\nu^2}\Ca+\landauca2\approx\left(\frac{1}{2}+1.146\varepsilon\right)\Ca+\landauca2
\label{eq:alpha1}
\end{equation}
where the numerical value of 1.146 corresponds to the sum of the first nine terms. The convergence of expression~\eqref{eq:alpha1} is demonstrated in \bild{fig:konvergenz}.
\begin{figure}
\centering%
\begin{tikzpicture}
\begin{axis}[
xmin=1,
xmax=9,
xtick={1,...,9},
ytick={1.15,1.2,...,1.4},
width=7cm,
height=5cm,
xlabel={number of terms},
ylabel=$(\alpha\ord1-1/2)/\varepsilon$,
]
\addplot[blue,mark=*] coordinates {(1,1.375) (2,1.18164) (3,1.1582) (4,1.15141) (5,1.14869) (6,1.14738) (7,1.14667) (8,1.14625) (9,1.14599)};
\end{axis}
\end{tikzpicture}
\caption{Convergence plot for the $\landaueps{}$-term in \gleich{eq:alpha1}.}%
\label{fig:konvergenz}%
\end{figure}
By means of the result for~$\alpha\ord1$ we can write down the final formula describing the interfacial shape
\begin{equation}
\tilde h\ord1(\tilde\varrho,\vartheta)=\sum_{\nu=1}^{\infty}\frac{\tau_\nu\ord0}{(1-4\nu^2)\tilde\varrho^{2\nu}}\cos\vartheta
\label{eq:hlsg2}
\end{equation}
which is visualized in \bild{fig:Partikel1} for a capillary number of~$\Ca=0.3$ and an equilibrium contact angle of~$\Theta=75^\circ$.
\begin{figure}
\centering%
\includegraphics[width=0.6\columnwidth]{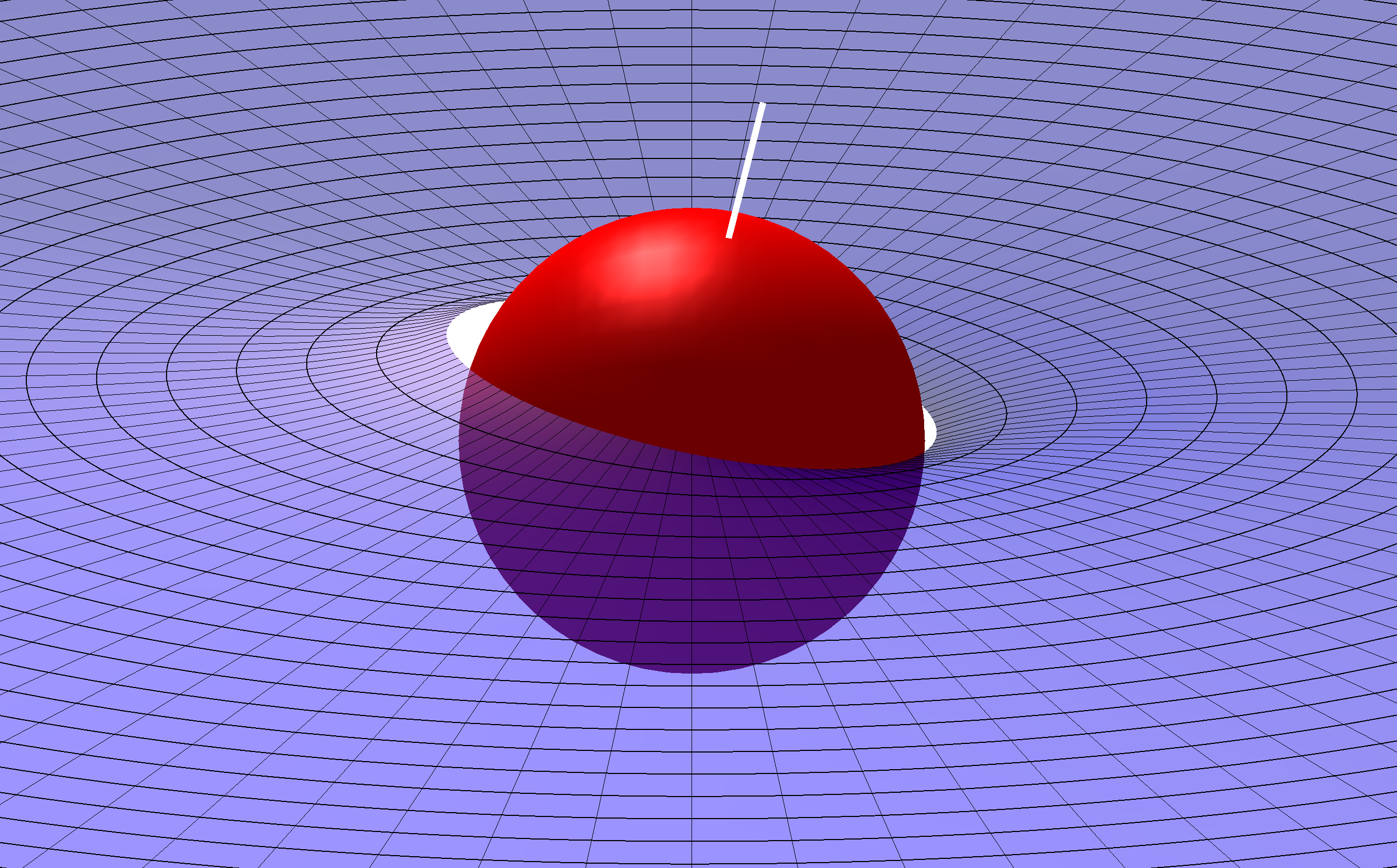}
\caption{Visualisation of the $\landauca1$-configuration for an equilibrium contact angle~$\Theta$ of~75\textdegree{} and a capillary number of~$\Ca=0.3$; the trim angle can be estimated from the inclination of the white line; the gap between the particle surface and the fluid-fluid interface arises because the boundary condition is evaluated on a cylindrical surface according to equation~\eqref{eq:hentwTCL}.}%
\label{fig:Partikel1}%
\end{figure}
The lower half-space is occupied by fluid~1 and the particle is moving to the right of the picture. A white line piercing through the north and south pole of the sphere has been added to indicate the change in orientation, that is, the trim angle. From \bild{fig:Partikel1} it can be seen that the contact angle changes around the particle's circumference. While the contact angle retains its equilibrium value at~$\vartheta=\pi/2$ and~$3\pi/2$, it deviates most strongly from that value at~$\vartheta=0$ and~$\pi$. From geometric considerations (see also \bild{fig:Ellipse}), the deviation at~$\vartheta=0$ is found to be
\begin{equation}
\delta\Theta=\sum_{\nu=1}^{\infty}\frac{\tau_\nu\ord0}{1-2\nu}\Ca+\landauca2\approx\left(\frac{3}{2}+2.863\varepsilon\right)\Ca+\landauca2,
\label{eq:dtheta}
\end{equation}
where again nine terms have been included in the summation. At~$\vartheta=\pi$, the contact angle difference~$\delta\Theta$ has the same magnitude but the opposite sign. Figure~\ref{fig:Schnitt} shows cross-sections of the interfacial shape and the TCL in the plane~$\tilde y=0$ for three different contact angles~$\Theta$.
The particle trim angle, indicated by the inclination of the TCL plane between~$\tilde z=-1$ and~$\tilde z=1$, obviously increases when the particle plunges deeper into fluid~1 (larger~$\varepsilon$) with its relatively high viscosity, which is an intuitive result.

\section{Pair interaction via Linear Superposition Approximation}

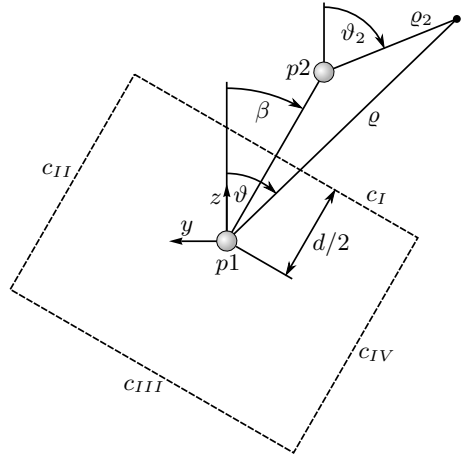
\begin{figure}
\centering%
\def\breitea{6cm}
\begin{tikzpicture}[x=0.06*\breitea,y=0.06*\breitea]
\node[anchor=south west,inner sep=0pt] at (0,0) {\includegraphics[width=\breitea]{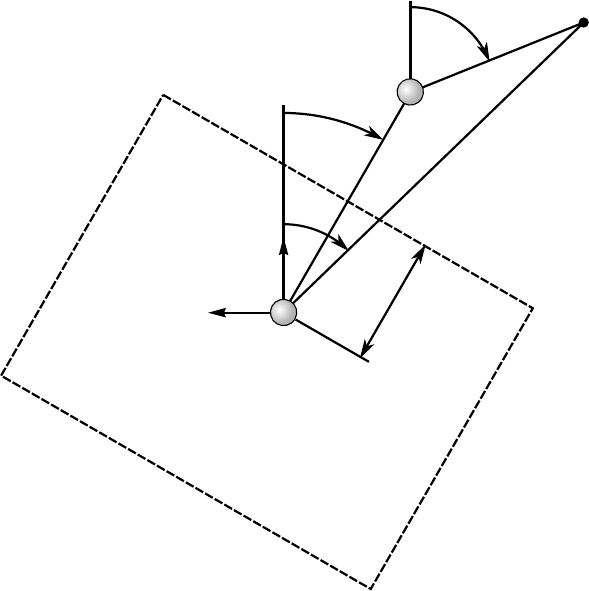}};
\coordinate (S1) at (8.1,7.9);
\path (S1) ++ (xyz polar cs:angle=75,radius=4.9) node {$\beta$};
\path (S1) ++ (xyz polar cs:angle=75,radius=1.8) node {$\vartheta$};
\path (S1) ++ (xyz polar cs:angle=40,radius=7) node {$\varrho$};
\path (S1) ++ (xyz polar cs:angle=270,radius=0.9) node {$p1$};
\path (S1) ++ (xyz polar cs:angle=-2,radius=3.8) node {$d/2$};
\path (S1) ++ (xyz polar cs:angle=107,radius=1.6) node {$z$};
\path (S1) ++ (xyz polar cs:angle=163,radius=1.6) node {$y$};
\coordinate (S2) at (11.7,14.2);
\path (S2) ++ (xyz polar cs:angle=28,radius=4) node {$\varrho_2$};
\path (S2) ++ (xyz polar cs:angle=52.5,radius=1.7) node {$\vartheta_2$};
\path (S2) ++ (xyz polar cs:angle=180,radius=1) node {$p2$};
%
\path (S1) ++ (5.5,1.6) node {$c_\Rmnum1$};
\path (S1) ++ (5.5,-4.2) node {$c_\Rmnum4$};
\path (S1) ++ (-6.3,2.6) node {$c_\Rmnum2$};
\path (S1) ++ (xyz polar cs:angle=-120,radius=6.2) node {$c_\Rmnum3$};
\end{tikzpicture}
\caption{Coordinate systems for two interacting particles~$p1$ and~$p2$ of separation~$d$ viewed from the negative $x/\chi$-direction; integration path~$c=c_\Rmnum1\cup c_{I\!I}\cup c_{I\!I\!I}\cup c_{I\hspace{-0.5pt}V}$ used in the evaluation of the capillary force.}%
\label{fig:Interaktion_Ksys}%
\end{figure}

Within the LSA, the interfacial deformation~\eqref{eq:hlsg2} around a single particle is used to construct the two-particle deformation necessary for studying capillary interactions via linear superposition. Since the method can only be expected to be accurate at large inter-particle separation~$d\ll a$, we truncate the infinite series in \gleich{eq:hlsg2} after the first, asymptotically dominant, $\tilde\varrho^{-2}$-term. Returning to dimensional quantities, but keeping the capillary number as an abbreviation, we may express the deformation~$h_{pi}$ around a single particle~$pi$ in the form
\begin{equation}
h_{pi}\approx-\tau_1\ord0\Ca_{pi}\frac{a^3}{3\varrho^2_i}\cos\vartheta_i=\Ca_{pi}\left(\frac{1}{2}+\frac{11}{8}\varepsilon\right)\frac{a^3}{\varrho^2_i}\cos\vartheta_i,\quad i=1,2.
\label{eq:hpi}
\end{equation}
In \gleich{eq:hpi}, the cylindrical coordinate system~$(\varrho_i,\vartheta_i,\chi)$ is centred at particle~$i$. Since we will later use only the coordinate system of particle~1, we set~$(\varrho_1,\vartheta_1,\chi)=(\varrho,\vartheta,\chi)$ for simplicity (see \bild{fig:Interaktion_Ksys}). The notation~$\Ca_{pi}$ instead of~$\Ca$ accounts for a potentially different velocity between the particles (in magnitude and sign). This may apply to the case when the particles differ in the properties governing their reaction to the external field driving them. Figure~\ref{fig:Interaktion_Ksys} shows the two particles with their centres separated by a distance~$d\gg a$ and with the line connecting their centres coinciding with the coordinate line~$\vartheta=\beta$.
In the course of the LSA, it is required to re-express the single-particle deformation~$h_{p2}$, which is given by \gleich{eq:hpi} within the~$(\varrho_2,\vartheta_2,\chi)$-system, in terms of the coordinates~$(\varrho,\vartheta,\chi)$. From \bild{fig:Interaktion_Ksys}, we can geometrically derive the relations
\begin{equation}
\label{eq:geo12}
\begin{split}
\varrho_2&=\sqrt{\varrho^2+d^2-2\varrho d \cos(\vartheta-\beta)},~\text{and}\\
\cos\vartheta_2&=\frac{\varrho\cos\vartheta-d\cos\beta}{\varrho_2},
\end{split}
\end{equation}
so that we have
\begin{equation}
h_{p2}\approx a^3\left(\frac{1}{2}+\frac{11}{8}\varepsilon\right)\Ca_{p2}\frac{\varrho\cos\vartheta-d\cos\beta}{\left[\varrho^2+d^2-2\varrho d \cos(\vartheta-\beta)\right]^{3/2}}.
\label{eq:hp2}
\end{equation}
Regarding the interfacial deformation~$h_\mathrm{LSA}$ in the presence of two particles, the LSA is based on the assumption that
\begin{equation}
h_\mathrm{LSA}=h_{p1}+h_{p2}.
\label{eq:hLSA}
\end{equation}
Physically, the dynamical deformation of the interface by the particles gives rise to an interaction force. In the present case of two moving particles, the interaction force is expected to be dependent on both the relative positions of the particles (described by~$\beta$), their separation, and the respective capillary numbers~$\Ca_{pi}$. To be sure, the boundary conditions on the particle surfaces are generally not fulfilled by the sum of the deformations according to \gleich{eq:hLSA}. However, in the region halfway between the particles, the approximation is supposed to be accurate when the separation~$d$ is large. Therefore, a reasonable path of integration ($c=c_I\cup c_{I\!I}\cup c_{I\!I\!I}\cup c_{I\hspace{-0.5pt}V}$, see \bild{fig:Interaktion_Ksys}) for the interfacial tension force runs through that region and is closed at infinite distance from the particles. The interaction force~$\vek F_\mathrm{int}$ may be written as
\begin{equation}
\vek F_\mathrm{int}=\oint\limits_c\sigma_{12}\left(\partiell{\vek r_c}{s}\times\vek n_{12}\right)\mathrm{d}s
\label{eq:Fint}
\end{equation}
where the parametrization of a point~$\vek r_c$ on the integration path~$c$ based on~$s$ is chosen such that the integrand becomes a vector pointing away from particle~1. Prior to its evaluation, the integral in \gleich{eq:Fint} can be simplified considerably with the aid of \bild{fig:Interaktion_Ksys}. First, we understand that the contributions along the paths~$c_{I\!I}$ and~$c_{I\hspace{-0.5pt}V}$ mutually cancel because the interfacial deformation vanishes at infinite distance from the particles. We are left with the integrals along~$c_I$ and~$c_{I\!I\!I}$ which diverge when evaluated separately since the interfacial tension force (more specifically, its line density) becomes constant at infinity. Therefore, both integrals need to be evaluated simultaneously. In effect, the contribution from~$c_{I\!I\!I}$ can be taken into account by subtraction of a vector of constant orientation under the integral since, again, the interfacial force is constant at infinity.  Choosing~$\mathrm{d}s=\mathrm{d}\vartheta$, the vector~$\vek r_c$ describing a point on the integration path is given by
\begin{equation}
\vek r_c=\frac{d}{2\cos(\vartheta-\beta)}\vek e_\varrho+\left.h_\mathrm{LSA}\right|_{\varrho=d/[2\cos(\vartheta-\beta)]}\vek e_\chi
\label{eq:cvek}
\end{equation}
with~$h_\mathrm{LSA}$ according to \gleich{eq:hLSA}. The normal vector, which is to be evaluated at~$\vek r_c$ using~$h_\mathrm{LSA}$, is calculated from \gleich{eq:n12lin}. As discussed above, we need to subtract a vector
\begin{equation}
\sigma_{12}\partiell{}{\vartheta}\left(\frac{d}{2\cos(\vartheta-\beta)}\vek e_\varrho\right)\times\vek e_\chi
\label{eq:cconst}
\end{equation}
representing the contribution from~$c_{I\!I\!I}$. Note that with the parametrization~$\mathrm{d}s=\mathrm{d}\vartheta$ the vector~\eqref{eq:cconst} is not constant as it would have been if~$\mathrm{d}s$ had been set equal to $\mathrm{d}y$, say. The foregoing expressions are valid for any choice of~$\Ca_{p1}$ and~$\Ca_{p2}$. Within the following discussion, however, we focus on the most relevant case of two particles having identical velocity vectors~$U\vek e_z$ and thus set~$\Ca=\Ca_{p1}=\Ca_{p2}$. As the evaluation of \gleich{eq:Fint} and of the analogous relation for the torque~$\vek M_\mathrm{int}$ with respect to the centre of particle~1 are straightforward from the above considerations, we may directly state the results
\begin{align}
\frac{\vek F_\mathrm{int}}{\sigma_{12}a}&=-16H\frac{a^2}{d^2}\cos\beta\vek e_x+3\pi H^2\frac{a^5}{d^5}\left\{\left[4+13\cos(2\beta)\right]\sin\beta\vek e_y+26\cos\beta\sin^2\beta\vek e_z\right\}\label{eq:Fxyz}\\%
\frac{\vek M_\mathrm{int}}{\sigma_{12}a^2}&=-\frac{51}{4}\pi H^2\frac{a^4}{d^4}\sin(2\beta)\vek e_x-4H\frac{a}{d}\left[2\cos^2\beta\vek e_y+\sin(2\beta)\vek e_z\right]\label{eq:Mxyz}
\end{align}
in~$(x,y,z)$-coordinates, where $H\definiert(1/2+11\varepsilon/8)\Ca$. Recall that the particle motion due to the external force is aligned with the~$z$-direction. The first term~$\sim d^{-2}$ in \gleich{eq:Fxyz} only results in vertical ($x$-)adjustment of the particle positions. On the other hand, the remaining two terms~$\sim d^{-5}$ act in the plane of the undisturbed fluid interface. In \bild{fig:Karte_int} the force components are shown qualitatively in the form of a vector pinned at a position determined by the respective value of~$\beta$.
\begin{figure}
\centering%
\def\breitea{5cm}
\begin{tikzpicture}[x=0.06*\breitea,y=0.06*\breitea]
\node[anchor=south west,inner sep=0pt] at (0,0) {\includegraphics[height=\breitea]{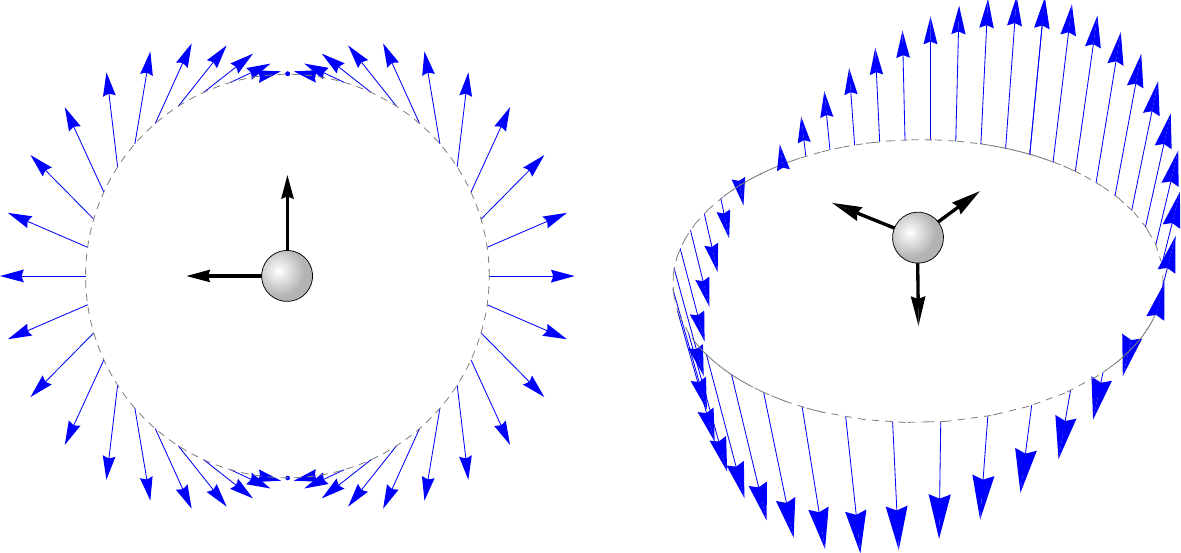}};
\coordinate (S1) at (8.7,8.3);
\path (S1) ++ (xyz polar cs:angle=76,radius=2.6) node {$z$};
\path (S1) ++ (xyz polar cs:angle=198,radius=2.6) node {$y$};
\path (S1) ++ (xyz polar cs:angle=-45,radius=1.4) node {$p1$};
\path (S1) ++ (xyz polar cs:angle=17,radius=6.1) node[circle, fill=black,inner sep=2pt] {};
\coordinate (S2) at (27.8,9.5);
\path (S2) ++ (xyz polar cs:angle=17,radius=1.7) node {$z$};
\path (S2) ++ (xyz polar cs:angle=137,radius=1.8) node {$y$};
\path (S2) ++ (xyz polar cs:angle=248,radius=1.9) node {$x$};
\path (S2) ++ (xyz polar cs:angle=-45,radius=1.4) node {$p1$};
\node at (1,15) {(a)};
\node at (21,15) {(b)};
\end{tikzpicture}
\caption{Interaction map qualitatively describing the components of the interaction force~\eqref{eq:Fxyz} acting on particle~1 dependent on the relative particle positions: (a) $y$- and~$z$-components, (b) $x$-component; the correspondence between length of the vectors and the force magnitude differs between the subfigures.}%
\label{fig:Karte_int}%
\end{figure}
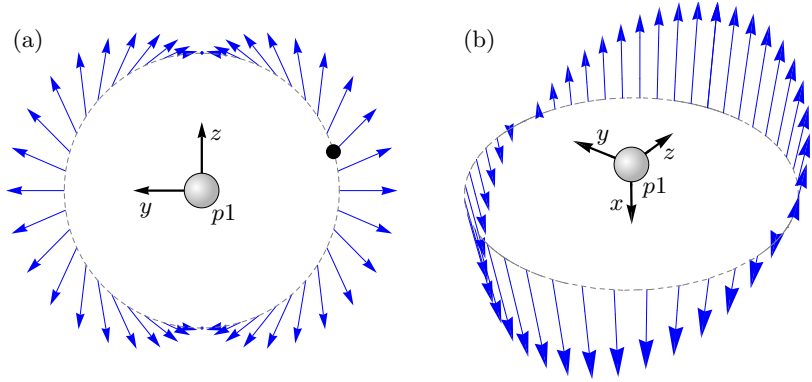
The interaction map may be read as follows. For a given~$\beta$, the connecting line between the particles~$p1$ and~$p2$ intersects the dashed circle in \bild{fig:Karte_int}. The vector plotted at the intersection point represents the force acting on particle~1 within the actual configuration. Figure~\ref{fig:Karte_int}(a) displays the~$y$- and~$z$-components of the interaction force while \bild{fig:Karte_int}(b) shows its~$x$-component. The scaling of the forces differs between figures~\ref{fig:Karte_int}(a) and~(b) for reasons of optimal visualization, owing to the different dependences of the force components on the separation~$d$ (see \gleich{eq:Fxyz}). In addition to the direction and the relative strength of the forces depending on the particle configuration, \bild{fig:Karte_int} also contains information on the torque resulting from the interaction. It can be shown that the forces sketched in the figure as pinned to the dashed circle do indeed provide a full representation of the force distribution, meaning that the vector between the centre of particle~$p1$ and the respective intersection point on the dashed circle, namely
\begin{equation}
-\frac{1}{2}d\sin\beta\vek e_y+\frac{1}{2}d\cos\beta\vek e_z
\label{eq:verbindung}
\end{equation}
(see \bild{fig:Interaktion_Ksys}), equals the lever arm for the calculation of the torque on the particle by means of the interaction force. In other words, the relation
\begin{equation}
\vek M_\mathrm{int}=\left(-\frac{1}{2}d\sin\beta\vek e_y+\frac{1}{2}d\cos\beta\vek e_z\right)\times\vek F_\mathrm{int}
\label{eq:FM}
\end{equation}
holds, which can be easily verified via the equations~\eqref{eq:Fxyz} and~\eqref{eq:Mxyz}. As an example, consider the black dot in \bild{fig:Karte_int} symbolizing an intersection point. The force on particle~1 can be read off as pointing to the northeast of the picture while the resulting torque tends to rotate the particle in the anticlockwise direction because the line of action of the force does not run through the particle centre but passes south of it.\par
From \gleich{eq:Fxyz} and, accordingly, \bild{fig:Karte_int} it can be seen that the interaction force vanishes when the particles move along a common line. When the second particle's position is on the $y$-axis, the force is attractive and acts along the line of centres. Between those extreme cases the interaction force changes continuously. The force component~$F_{||}$ along the line of centres
\begin{equation}
F_{||}\definiert\vek F_\mathrm{int}\cdot(-\sin\beta\vek e_y+\cos\beta \vek e_z)=\frac{27\pi}{2d^5}H^2\left[1-\cos(2\beta)\right]\geq0
\label{eq:Fpar}
\end{equation}
is generally attractive. Following the commonly accepted method of analysing interfacial deformation by means of a multipole expansion (\citealp[equation (2.16)]{Danov2005}; \citealp{Oettel2008}), we identify the interfacial shape~\eqref{eq:hlsg2} as dipolar. As a consequence, the interaction between particles driven by an external force, given by \gleich{eq:Fxyz}, may also be termed dipolar. In fact, the interaction force between two fixed electric dipoles with equal dipole moments lying within a common plane has the form \citep[table~2.2, setting~$\theta_1=\theta_2=\pi/2-\beta$ and~$\phi=0$]{Israelachvili2011}
\begin{equation}
F_{||,\mathrm{el}}\sim -1+3\cos(2\beta)
\label{eq:Fel}
\end{equation}   
showing the same variation with~$\beta$ as in the case of driven particles.

Since the particles move along the fluid-fluid interface and their motion causes fluid flow, the discussion of interactions between the particles is only complete if we comment on hydrodynamic interactions. For two equal spheres at low Reynolds number in an unbounded fluid acted upon by the same force, it is well known that the particle velocity vectors are always parallel (\citealp[][p. 52]{Russel1989}; \citealp[][pp. 242-243]{Happel1983}). In that context, the particles' direction of motion need not be parallel to the driving force vector and the inter-particle separation remains constant during the motion. In the case of particles moving along a fluid-fluid interface considered in the present work, we have already seen that the hydrodynamic problem at zero capillary number can be treated by means of an equivalent problem of a particle in an unbounded fluid (cf. \bild{fig:symm}). Therefore, the aspect of hydrodynamic interactions for the $\landauca0$-problem is also equivalent to the interaction of particles in an unbounded fluid. The same reasoning has been followed by \citet{Vassileva2005}. As a consequence, the hydrodynamic interaction between particles moving along a fluid interface by action of the same external driving force does not influence their separation, that is, the hydrodynamic force vectors acting on each particle are equal. We conclude that capillary effects govern the interaction between driven interfacial particles at least up to order~$\Ca$.

In order to estimate the strength of the dipolar interaction force, varying as~$d^{-5}$ according to \gleich{eq:Fxyz}, a comparison with another capillary interaction force is instructive, namely the quadrupolar force arising from an undulated three-phase contact line \citep{Stamou2000,Danov2005,Dominguez2008}. The latter also decays as $d^{-5}$. Using the values given by \citet{Stamou2000} in their estimation, we arrive at the quadrupolar interaction force~$F_\text{\ding{67}}= 48\pi\sigma_{12}a^{-1} \delta x\, \tilde d^{-5}$. Thereby,~$\delta x$ is a characteristic undulation of the TCL. \citet{Park2011} report a value of~$\delta x/a=0.03$ for the particles investigated, leading to~$F_\text{\ding{67}}\approx 0.14\,\sigma_{12}a \tilde d^{-5}$. On the other hand, the parallel component of the dipolar interaction force in the case~$\varepsilon=0$ and~$\beta=\pi/2$ is~$F_{||}\approx 21.21\sigma_{12}a \Ca^2 \tilde d^{-5}$. Therefore, the two forces are equal at a capillary number of~$\Ca\approx0.08$ corresponding to a particle velocity of~$U\approx1.7\,$mm/s for the 1000\,cSt silicone oil used by \citet{Ally2010}. If, of course, the three-phase contact line is less irregular, the dipolar force may dominate at an even smaller velocity. The same obviously holds for a liquid of higher viscosity.

\section{Conclusions}

In this work, we have studied the problem of spherical particles driven along a fluid-fluid interface by an external force. By simultaneously using two asymptotic expansions, we have succeeded in calculating the flow field for equilibrium contact angles not too far away from~90\textdegree{} and the resulting fluid-fluid interfacial deformation. The three-phase contact line at the surface of the particle has been assumed to be pinned, implying a variation in the contact angle along the surface. At the same time, the particle's response to the torque resulting from the asymmetric drag associated with the difference in fluid viscosity across the fluid-fluid interface has been described. More specifically, the particle rotates by a certain angle which we have termed the trim angle. Knowing the interfacial deformation around a single particle, we have described the capillary interaction between two particles through the Linear Superposition Approximation. The interaction force has been shown to be dipolar in terms of the azimuthal angle and to decay with the fifth power of the separation, similar to a capillary quadrupole arising from undulations of the contact line. To visualize the interaction force vector for varying configurations of the particles we have introduced an interaction map from which the torque acting on a particle along with the interaction force can be deduced. Hydrodynamic interactions have been discussed and shown to be negligible. The strength of the dipolar interaction has been compared to interactions resulting from undulated contact lines, showing that for particles moving sufficiently fast, the dipolar interaction is dominant.

The reasoning followed in the present work can be applied to particles of more general shapes than spheres, provided that the undeformed, equilibrium fluid-fluid interface is planar, which allows for the construction of an equivalent flow problem by reflection (as shown in \bild{fig:symm}). On the methodological side, corrections to the asymptotic formalism developed by \citet{Brenner1964} are required, as has been demonstrated in section~\ref{page:Brenner}. Furthermore, our analysis can easily be extended to incorporate a Navier slip condition on the particle surface, which may become relevant for nanoparticles.

As discussed above, the particle motion is associated with a configurational change parametrized by the trim angle. When the corresponding change in contact angle exceeds the range of contact angles permitted by contact angle hysteresis, we expect a qualitative change of the situation. For instance, the particle may detach from the interface and cause anomalous diffusional behaviour as reported by \citet{Chen2008} and \citet{Sriram2012}.

The above analysis is based on the assumption of small Reynolds numbers leading, in particular, to a flow field with a fore-and-aft symmetry. Consequently, we have found an interfacial shape displaying the same type of symmetry. For finite Reynolds numbers, however, the interfacial shape becomes asymmetric and evolves into a fascinating wave pattern \citep{Moisy2014} implying additional wave-making resistance. A possible route to the analytical treatment of the problem, which is left for future research, could rely on the Oseen approximation. 

\bibliographystyle{jfm}

\bibliography{LiteratureDrivenParticle}

\end{document}